# Why Do We Suffer for Fun? Ordeal Pleasure in Souls-like Games

**Flint Xiaofeng Fan**
ETH Zurich
Zurich, Switzerland
fxf@u.nus.edu

## ABSTRACT

Souls-like games exemplify how digital play can produce radical forms of pleasure through sustained challenge: players voluntarily invest tens or hundreds of hours in experiences designed to kill them repeatedly. This paper theorizes ordeal pleasure as a community-level phenomenon that emerges most fully when three mechanisms reinforce one another: Ludic Cultivation (mastery through fair adversity), Aspirational Deferment (delayed gratification oriented toward future growth), and Communal Mythopoesis (collective construction of shared meaning). Drawing on game design analysis, empirical player studies, and community discourse, we show how Souls-like games produce pleasure by coordinating difficulty, temporal structure, and social meaning-making. Comparative analysis (Elden Ring, Hollow Knight, Lords of the Fallen, The Surge) illustrates how specific design choices enable or undermine ordeal pleasure. The framework adds a temporal dimension to motivation theory and specifies social mechanisms beyond generic relatedness. It also offers design principles for designers seeking to create challenging games that transform suffering into complex satisfaction.

## INTRODUCTION

A characteristic gameplay pattern in FromSoftware's Elden Ring (2022) involves players attempting Malenia, Blade of Miquella, late into the night despite work the next morning and dozens of easier games in their library. Players fix their eyes on attack patterns, learning the rhythm of the Waterfowl Dance through repeated deaths. Qualitative studies of Souls-related community discussions record repeated attempts to articulate this persistence and to describe the pull of continuing through struggle (e.g., Czauderna et al. 2024; Väkevä et al. 2025). Nor is the broader appeal of this kind of challenge marginal: Elden Ring surpassed 25 million copies by mid-2024 and 30 million by April 2025 (Bandai Namco Entertainment 2025). Players voluntarily spend tens of hours in games explicitly designed to kill them repeatedly.

This presents a fundamental paradox: Why do players voluntarily choose suffering? Why is this experience pleasurable rather than simply frustrating? This is not straightforward masochism in the clinical sense. Rozin et al. (2013) use benign masochism to describe the enjoyment of aversive experiences under conditions of perceived safety and control, a framing that helps clarify why difficult play need not collapse into pathology. Qualitative work on Dark Souls communities likewise documents how players describe struggle as meaningful and sometimes beneficial (e.g., Väkevä et al. 2025). From the design side, Wilson and Sicart's (2010) notion of abusive game design helps explain how antagonistic systems can be structured to cultivate such experiences.

These perspectives help explain why aversive play need not collapse into pathology or arbitrary frustration. They also connect to work on emotional challenge in games (e.g., Bopp et al. 2016; Bopp et al. 2018), but our focus is on the structures that make such challenge sustainable in Souls-likes—specifically, how difficulty, motivation, learning, and community reinforce one another rather than merely co-occurring.

**Flow theory** (Csikszentmihalyi 1990; Chen 2007) predicts optimal experience when challenge and skill are balanced within the flow channel (see Figure 1). Souls games, however, ask players to remain outside that equilibrium for extended periods. Petralito et al. (2017), in a study of Dark Souls III players (N = 95), found that achievement experiences were overwhelmingly linked to difficulty and failure yet still rated highly for enjoyment. Flow theory therefore helps name the balance it expects, but not why players persist through prolonged frustration or why that frustration can itself become satisfying.

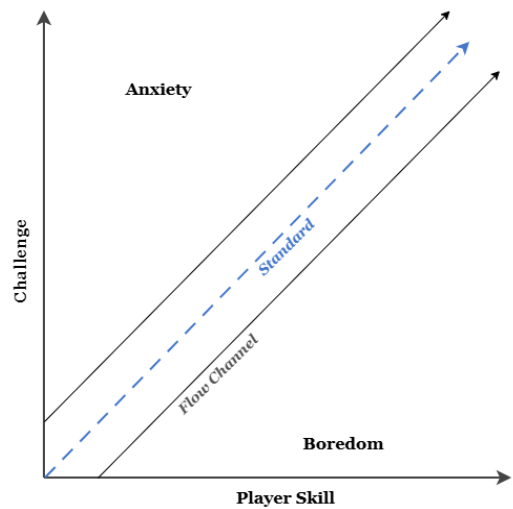

**Figure 1: The Flow Channel.** The band represents Csikszentmihalyi's (1990) flow channel, where challenge and skill are roughly balanced. The standard design trajectory (dashed line) keeps players within this zone, scaling difficulty in proportion to growing skill. Above the channel lies the anxiety zone (challenge exceeds skill); below lies boredom (skill exceeds challenge). Flow theory predicts that optimal engagement occurs within this band—a present-moment equilibrium that does not account for the extended periods of deliberate imbalance characteristic of Souls-like design (see Figure 2).

**Self-Determination Theory** (Ryan, Rigby, and Przybylski 2006) explains game motivation through competence, autonomy, and relatedness. Souls games clearly engage all three, yet SDT still leaves unresolved why frustration can amplify later satisfaction and why Souls-like communities sustain unusually intense collaborative interpretation alongside strategy sharing.

**Learning and mastery literature** (Bjork 1994; Bjork and Bjork 2011; Shute et al. 2015) helps explain how difficult systems cultivate persistence and long-term skill. Souls games exemplify this pedagogy through death-as-learning mechanics. Yet learning theory does not explain why this process is intrinsically enjoyable or why similar difficulty structures in other games often fail to engage players.

No existing framework explains how difficulty, motivation, learning, and community reinforce one another in ordeal pleasure. Existing theories treat them as separate phenomena that co-occur rather than as interacting mechanisms.

This paper applies a retrospective theoretical synthesis to the formative classic era of the Souls-like genre (2011-2022), spanning from Dark Souls to Elden Ring's initial release in 2022. We employ an abductive synthesis approach, integrating game design analysis of these foundational titles with secondary synthesis of empirical player studies. While we reference later scholarship (e.g., Czauderna et al. 2024; Väkevä et al. 2025), the core corpus centers on this classic period, and those later studies primarily analyze datasets rooted in it, such as Dark Souls and Elden Ring Reddit discussions and player accounts from 2011-2022. Later updates and post-2022 titles are treated, when mentioned, only as brief illustrative cases rather than as foundations of the framework. This methodological choice keeps the model grounded in the genre's established core.

## Contribution and Structure

The Ordeal Pleasure Framework (OPF) describes a community-wide phenomenon rather than a universal one: players vary substantially in which mechanisms they engage and in how strongly they do so. Nguyen's (2020) distinction between "striving players," who value challenge and skill development intrinsically, and "achievement players," who prioritize completion, suggests that ordeal pleasure in its fullest form may be more characteristic of the former. Even within this orientation, players differ in frustration tolerance, available time, and social inclination. Some engage deeply with communal lore; others actively avoid it. The framework therefore specifies the structural conditions under which ordeal pleasure becomes possible and more intense, rather than positing a mechanism that functions identically for all players.

This framework is developed through abductive theoretical synthesis (Dubois and Gadde 2002), moving iteratively between theoretical propositions and three bodies of material. First, we analyze the design of Dark Souls, Elden Ring, and comparison titles, focusing on how difficulty, progression, and narrative delivery are structured. Second, we synthesize findings from existing empirical player studies (Petralito et al. 2017; Väkevä et al. 2025; Shute et al. 2015). Third, we draw selectively on published discourse analyses and illustrative public community materials, especially the Dark Souls and Elden Ring discussion corpora analyzed by Czauderna et al. (2024) and Väkevä et al. (2025). This approach brings structural, psychological, and cultural evidence into dialogue, but it does not claim original large-scale discourse coding.

The paper proceeds as follows. We begin by examining existing theories, identifying what each illuminates and what each leaves unaccounted for in order to show why an integrative account is warranted. We then present the Ordeal Pleasure Framework, defining each mechanism with theoretical grounding and empirical support before examining their interaction. Next, we apply the framework through comparative case analysis, using the cases illustratively rather than as decisive validation. The paper concludes with the framework's theoretical and design implications.

## BACKGROUND

The introduction has identified the explanatory gap; this section clarifies how the main theoretical traditions illuminate parts of the phenomenon while leaving key aspects unresolved. Juul's (2013) paradox of failure identifies the central tension of enjoying negative affect, while Wilson and Sicart (2010) describe abusive game design in which designer-player antagonism creates meaning. Drawing on Daniel Vella's (2015) concept of the ludic sublime—the aesthetic power arising from the tension between the player's striving for mastery and systemic unknowability—we map our three mechanisms as follows: Ludic Cultivation captures the mastery side of this tension, Aspirational Deferment the temporal experience of dwelling in the gap between present inability and anticipated mastery, and Communal Mythopoesis the collective work of turning systemic opacity into shared meaning. We also build on the emotional challenge literature (Bopp et al. 2016; Bopp et al. 2018), which helps situate the negative affect at stake here and shows how it can be sustained over tens or hundreds of hours.

### Flow Theory: The Temporal Blind Spot

Flow theory (Csikszentmihalyi 1990; Chen 2007) posits that optimal engagement occurs when challenge matches skill. However, this model does not account well for the frustration-persistence paradox in Souls-likes, where players willingly endure extended periods in the anxiety zone (challenge > skill). Petralito et al. (2017) found that achievement experiences in Dark Souls III were overwhelmingly linked to failure, yet rated highly for enjoyment. Some players may indeed endure frustration instrumentally on the way to eventual flow. Even so, flow theory does not fully explain why the frustration phase itself can feel worthwhile, nor how later mastery can retrospectively reframe that frustration. The critical limitation here is temporal: flow emphasizes present-moment balance, whereas Souls players endure present imbalance for future mastery (see Figure 2). This ordeal trajectory suggests that players are not merely waiting for flow but investing in Aspirational Deferment.

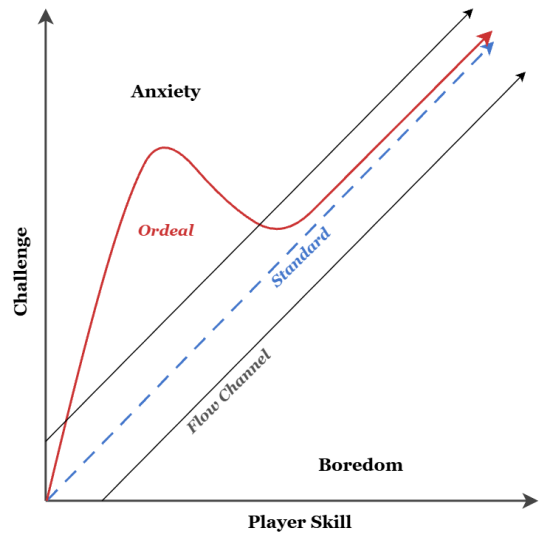

**Figure 2: The Ordeal Trajectory.** The shaded band marks the flow channel. The standard design trajectory remains within this safe zone, while the Souls-like trajectory deliberately pushes players into an extended "anxiety" phase. This sustained imbalance functions not as failure, but as an investment in future mastery, converting temporary frustration into the intensified satisfaction of "ordeal pleasure."

## Self-Determination Theory: The Competence Frustration Paradox

While Self-Determination Theory (Ryan, Rigby, and Przybylski 2006) explains motivation via the needs for competence, autonomy, and relatedness, it faces a competence frustration paradox in this genre. Standard SDT predicts that repeated failure should reduce enjoyment. Yet in Souls-likes, the period of frustration appears to amplify the eventual satisfaction, a dynamic recently explored as the compensation effect (Meng et al. 2024), though typically framed as behavioral repair rather than affective transformation. Separately, SDT's generic notion of relatedness does not capture the specific hermeneutic nature of Souls communities, which engage in interpretive labor that SDT's relatedness construct was not designed to address.

## Learning Theory: The Intrinsic Gap

Bjork's (1994) framework of desirable difficulties effectively describes the mechanics of Souls-like learning (spacing, interleaving, retrieval practice). The temporal and iterative dimension of gameplay engagement has also been theorized in game studies: Arsenault and Perron's (2008) magic cycle model conceptualizes player engagement as a process of building ever-greater knowledge across multiple interlinked interpretive dimensions over time, a framework that resonates with the progression structure we describe. However, learning theory treats satisfaction as a byproduct of skill acquisition, failing to explain why players find this specific form of pedagogical punishment intrinsically pleasurable when similar difficulties in other games lead to churn. It explains how players learn, but not why they endure the difficulty. This gap motivates our first mechanism, Ludic Cultivation, which reframes this pedagogical structure as an aesthetic rather than purely instrumental experience.

## The Integration Imperative

Taken together, these gaps motivate an integrative account. The Ordeal Pleasure Framework brings these theories into one account by arguing that Ludic Cultivation, Aspirational Deferment, and Communal Mythopoesis can reinforce one another in ways no single theory captures.

## THE ORDEAL PLEASURE FRAMEWORK

The Ordeal Pleasure Framework proposes three mechanisms that can reinforce one another in transforming challenging gameplay into meaningful satisfaction. We describe this process through a "Forging Ritual" metaphor, in which the player is both the smith and the metal.

- Ludic Cultivation (The Anvil & Hammer): Mastery through iterative engagement with fair, learnable adversity. The anvil must be solid (fair rules) for the strike to shape the metal rather than shatter it.
- Aspirational Deferment (Enduring the Heat): Delayed gratification oriented toward future growth. This is not passive waiting, but the active endurance of high temperatures (frustration) necessary for transformation.
- Communal Mythopoesis (The Legend): Collective construction of shared narrative and meaning. The story of the weapon's creation—the shared struggle—gives meaning to the heat and the hammer, transforming a mere tool into a legendary artifact.

Together, the three mechanisms specify how the ludic sublime is sustained in practice: Cultivation structures resistance, Deferment structures temporality, and Mythopoesis structures communal meaning. These mechanisms can reinforce one another, though Figure 3 presents this interaction as a heuristic rather than a demonstrated causal system. Comparative analysis suggests that weakening one mechanism can change the intensity or character of the fuller, community-wide form of ordeal pleasure.

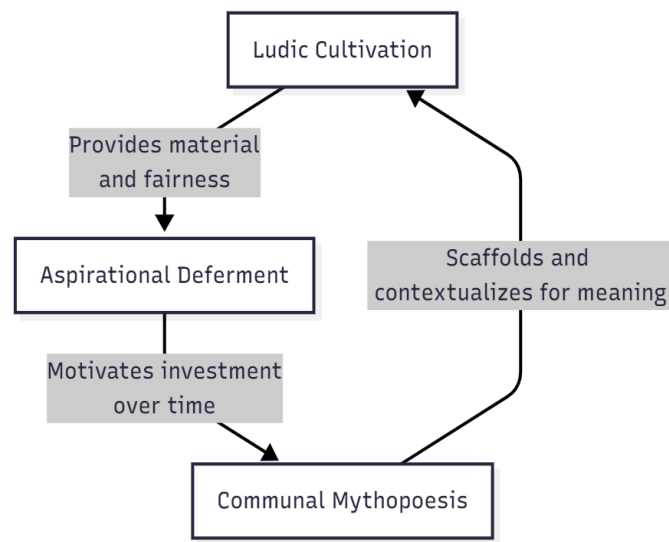

**Figure 3: The Ordeal Pleasure Framework.** Ludic Cultivation, Aspirational Deferment, and Communal Mythopoesis form a mutually reinforcing cycle: fair, learnable adversity provides material for mastery-

oriented goals; long-term temporal investment motivates communal participation; and collective meaning-making scaffolds and contextualizes individual cultivation.

## Ludic Cultivation: Mastery Through Structured Adversity

Ludic Cultivation is mastery development through iterative engagement with fair, learnable adversity. Drawing on Aarseth's (1997) account of ergodic literature, Souls games demand active, skill-dependent engagement rather than passive consumption. This connects to Bjork's (1994) notion of desirable difficulties but adds game-specific conditions. For cultivation to occur, difficulty must be fair, with readable telegraphing, consistent rules, and multiple viable approaches, and it must be iterative, so that failure teaches and visible progression eventually makes earlier obstacles trivial. In this configuration, death functions pedagogically rather than punitively.

### *The Mechanisms of Desirable Difficulty in Souls Design*

Souls games implement specific desirable difficulties that optimize long-term skill retention:

**Spacing and the Bonfire Mechanism:** The "spacing effect" (Bjork 1994) suggests that learning is greater when practice is distributed over time rather than massed in a single session. Bjork (1994) notes that when information is easily accessible (high retrieval strength), restudying provides little benefit. However, when retrieval strength decreases through spacing, the subsequent retrieval effort significantly boosts storage strength.

Souls checkpoint systems (Bonfires in Dark Souls, Sites of Grace in Elden Ring) can function as spacing mechanisms. When players die, they return to the last checkpoint and must traverse the environment again. That traversal may act as a spacing interval, giving retrieval strength time to decay slightly before the next attempt on the specific encounter. When players reach the enemy again, they must retrieve attack patterns from memory. This repeated retrieval effort may help explain why combat patterns become deeply learned over time, even if the precise transfer from laboratory spacing effects to action-game play should be treated cautiously.

**Interleaving vs. Blocking**: Bjork's (1994) research shows that while "blocking" (practicing one skill repeatedly: AAAA) leads to better immediate performance and higher confidence, "interleaving" (mixing skills: ABCABC) produces far superior long-term retention and transfer.

Souls combat is inherently interleaved. Players cannot practice "dodging" in isolation—they must dodge an arrow, then block a sword swing, then manage stamina, then heal, then position for counterattack. The game forces constant switching among different motor and attentional demands. This can feel cognitively taxing and can create the sensation that one is not improving, even as it contributes to more flexible skill acquisition. The struggle, on this reading, is not straightforward evidence of non-learning; it may be part of how learning is being distributed across multiple tasks at once.

**Variation and Context**: Learning theory also highlights encoding variability—learning in different contexts improves recall. Souls games frequently reuse enemies in different environments (e.g., the Capra Demon as a boss in a cramped room in Dark Souls, then later as a standard enemy in lava ruins). This forces players to adapt learned skills to new spatial contexts, further reinforcing storage strength of combat mechanics. The variation prevents rote memorization and encourages genuine pattern recognition.

Väkevä et al.'s (2025) CHI study on Dark Souls and mental health reports that players frame the game as teaching persistence and endurance. Welsh's (2020) theoretical analysis documents how mastery cultivation becomes embedded in community practice, from speedrunners to casual players.

Cultivation helps address flow theory's temporal blind spot. By reframing death as pedagogy rather than punishment, failure can become usable information. This creates what Lazzaro (2004) terms 'Hard Fun' and aligns with prior work on emotional challenge (Bopp et al. 2018). Design can make frustrating repetition function as distributed practice rather than wasted effort.

## Aspirational Deferment: Growth Through Delayed Gratification

Aspirational Deferment is the motivational orientation in which players willingly endure present frustration as an investment in future mastery. We distinguish this from the classic Marshmallow Test model of delayed gratification (Mischel et al. 1972), in which the subject simply waits for a larger reward later. In Souls-like play, players do not merely wait; they work through aversive states in order to transform their own capacities. It is closer to athletic training: players endure the heat of exertion not because waiting yields a prize, but because the process itself is transformative. Drawing on Ryan and Deci's (2000) account of competence and Dweck's (2006) growth mindset theory, this mechanism turns immediate difficulty into a long-term developmental goal, especially when games provide long horizons, visible progression, "come back later" affordances, and a design logic in which failure teaches rather than merely punishes.

This mechanism helps explain why frustration and achievement are so tightly entangled in Petralito et al.'s (2017) findings: negative experiences are recast as investments in future positive outcomes. Similarly, Väkevä et al. (2025) report that players articulate a growth-oriented understanding of failure and connect this mindset to real-world challenges, including depression management.

Deferment helps explain the frustration-satisfaction transformation that Self-Determination Theory leaves underspecified. Present negative affect becomes meaningful when embedded in a growth-oriented temporal structure. Players' qualitative accounts suggest that the depth of prior frustration can sharpen the later sense of mastery, consistent with contrast-based interpretations of affect (Meng et al. 2024). What distinguishes this mechanism from simple persistence is that players do not merely tolerate difficulty; they reinterpret present frustration as evidence of future growth. That temporal reframing also helps explain how prolonged struggle can become socially meaningful beyond the individual player, a point that becomes central to Communal Mythopoesis.

## Communal Mythopoesis: Collective Meaning-Making

Communal Mythopoesis is the collective construction of shared narrative, meaning, and cultural memory through collaborative interpretation and social practice. By withholding full explanations, Souls-like design can make complete understanding difficult in isolation and invite collaborative investigation. This differs from generic strategy sharing or routine wiki documentation. Strategy guides help players win; mythopoesis is the shared work of deciding what the world means and why it matters. Two modes matter here. The first is shared ordeal bonding: commiseration, solidarity, and collective identity arising when players face similar challenges together. The second is lore hermeneutics: the interpretive labor of reconstructing cryptic narrative fragments through discussion, videos, and wiki debate. While analytically distinct, these modes often overlap in Souls-like play: shared difficulty makes the ordeal worth narrating, and narrative gives the ordeal meaning beyond the mechanical fact of persistence.

Communal Mythopoesis depends on a sufficiently rich lore substrate, interpretive gaps that reward inference, memorable cultural touchstones, communication systems such as messages, bloodstains, and phantoms, and difficulty moments vivid enough to become collective points of reference.

These conditions do not merely make interpretation possible; in successful Souls-like communities, they can scale into substantial collective labor. Czauderna et al.'s (2024) discourse analysis found a single Elden Ring Reddit thread with 2,363 comments totaling 116,911 words. Multiple academic studies likewise note how cryptic lore generates substantial interpretive labor. Welsh's (2020) theoretical analysis shows how late-stage Dark Souls play develops distinctive community practices around collective mastery. Perreault and Lynch (2022) argue that gaming communities function as interpretive communities in Fish's (1980) sense: Souls players develop sophisticated hermeneutic practices that treat the game world as worthy of sustained analysis.

### *The Psychological Function of Communal Mythopoesis*

Recent research also shows that Communal Mythopoesis serves psychological functions beyond interpretive labor, especially through shared ordeal bonding. Väkevä et al.'s (2025) CHI study on Dark Souls and mental health identifies several ways community engagement can support players.

At the level of shared ordeal bonding, the game's rhythms can give players something clear to return to after failure, while the community may supply encouragement. In Väkevä et al.'s material, players describe the in-game phrase "Don't you dare go Hollow," later adopted by the community as a shared mantra, as motivational both in-game and in everyday life.

The game's symbolic representation of struggle—the curse, the dying world, the repeated death and resurrection—can also allow players to externalize and process their own difficulties (Väkevä et al. 2025). The community can provide space to discuss these symbolic connections, treating the game as a shared text through which personal challenges become discussable.

The community also often supports struggling players by using the game's difficulty as a shared language of empathy. Czauderna et al.'s (2024) CASEL-based analysis found Elden Ring Reddit discussions rich in self-management and relationship-skills language. Players vent frustration, and other users often respond with validation and encouragement; the upvoting system and shared lexicon can help make such exchanges more visible and more welcoming.

*The Ludic Representation of Social Support*

The game's multiplayer mechanics can be read as ludic representations of this shared-ordeal bonding. The "Summoning" mechanic—calling another player for help—makes the psychological dimension concrete. Players are alone in their world but can see "ghosts" of other players, see bloodstains where others have died, and read their messages. The message system enables asynchronous communication, with players leaving warnings ("Tough enemy ahead"), jokes ("Try finger, but hole"), or encouragement ("Don't give up, skeleton!") as illustrated in Figure 4.

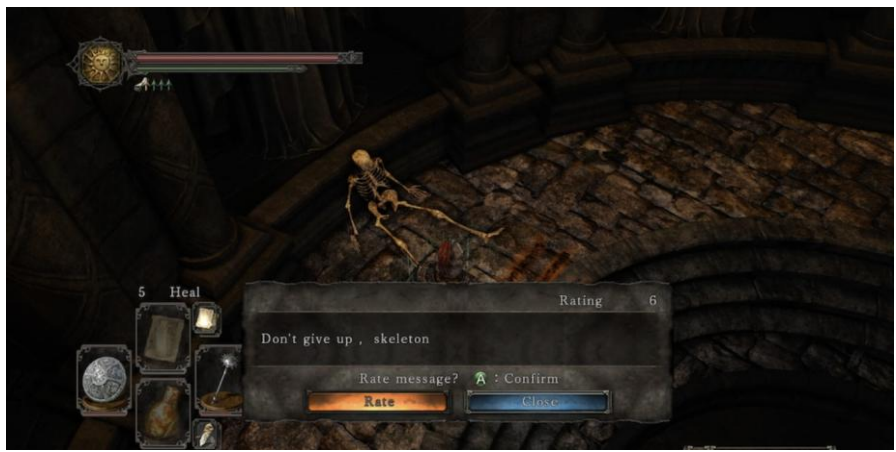

**Figure 4: Screenshot from Dark Souls II (FromSoftware 2014) showing the asynchronous message mechanic.** Players can leave text messages for others at specific locations, which appear as glowing signs in other players' worlds. This example shows the community-created encouragement "Don't give up, skeleton"—a player-authored message left next to a skeleton. This design transforms individual struggle into visible collective experience, as players encounter evidence of others' presence and solidarity without requiring synchronous multiplayer. The message system exemplifies how FromSoftware integrates Communal Mythopoesis into gameplay, making social support a ludic rather than purely social phenomenon.

This design can create a sense of shared ordeal: the visibility of other players' struggles can help normalize failure and create solidarity, and summoning help may provide both mechanical assistance and psychological validation.

Mythopoesis also helps address learning theory's relative blindness to intrinsic enjoyment. Individual suffering can become part of a shared legend, especially when players understand themselves as struggling alongside a much larger community. Interpretive labor gives difficulty narrative significance, and community support can transform isolated frustration into shared ordeal.

## Co-Constitution: How Mechanisms Interact

Evidence suggests these mechanisms reinforce one another within a cycle. Ludic Cultivation provides the foundation: fair, learnable adversity creates a believable path from incompetence to mastery, validating the growth mindset required for Aspirational Deferment. Without this 'solid anvil,' players may struggle to develop the belief in future competence that sustains them through failure. In turn, Aspirational Deferment gives the ordeal temporal meaning, turning extended struggle into an experience that can fuel Communal Mythopoesis. This temporal investment makes interpretive labor feel worthwhile; without it, community interaction remains more transactional. Finally, Communal Mythopoesis completes the cycle by scaffolding individual practice. The legend surrounding challenges such as Malenia means that players return already entering a shared story, not just another boss fight.

This synergy is illustrated by existing studies rather than directly verified as a full triadic mechanism. Petralito et al. (2017) link achievement to repeated failure; Welsh (2020) and Väkevä et al. (2025) describe community practices that send players back into cultivation with renewed growth orientations; and Czauderna et al. (2024) show how mechanical difficulty, affective support, and communal processing can reinforce one another within the same discussion spaces.

This account yields a comparative prediction: games that weaken any one mechanism may struggle to sustain the full-intensity, community-wide form of ordeal pleasure even when the others remain present. The next section tests that prediction through comparative analysis.

## COMPARATIVE ANALYSIS: PRESENCE AND ABSENCE OF MECHANISMS

The framework's explanatory power is illustrated through four deliberately contrasting cases rather than a representative sample: a broad-appeal Souls case (Elden Ring), an adjacent high-engagement difficult game (Hollow Knight), and two comparatively weaker or more partial Souls-like implementations (Lords of the Fallen and The Surge).

### High Integration, Permissive Structure: Elden Ring

Elden Ring brings the three mechanisms together, but does so through a more permissive structure than earlier Souls titles. It helps explain part of the genre's broad appeal without serving as the most tightly integrated case of the framework.

The game links Ludic Cultivation and Aspirational Deferment through its open-world structure. Players who hit a wall can explore elsewhere, gather tools, and return later, which can preserve challenge without forcing immediate abandonment. At the same time, that same structure can permit over-leveling that bypasses cultivation rather than completing it. Elden Ring is therefore best treated as a game that can support strong ordeal pleasure, rather than one that enforces it as tightly as Dark Souls or Bloodborne. This aligns with Felczak's (2025) reading of Elden Ring as broadening access to difficulty rather than demanding strict mastery in every case.

Elden Ring also supports intense communal mythopoesis. George R.R. Martin's public association with the project heightened expectations of narrative depth, while FromSoftware's cryptic presentation preserved the interpretive gaps regardless of the exact division of creative labor. Large-scale Reddit theory-crafting and cultural touchstones such as "Let me solo her" show how communal meaning-making can feed back into individual persistence. Difficulty generates shared stories, and those stories may make return more compelling for some players.

## Partial Integration: Hollow Knight as Instructive Contrast

Hollow Knight (Team Cherry, 2017) illustrates a different configuration: a critically acclaimed difficult game with strong cultivation and deferment.

Its Communal Mythopoesis appears less tightly coupled to difficulty than in Souls-likes. Hollow Knight has rich lore and an active theory community, but more of the narrative is discoverable through solo play, so collective interpretation seems less necessary. Hollow Knight therefore shows that cultivation and deferment can sustain strong engagement without the same degree of community intensity characteristic of Souls-like cultures. In designs where isolation is itself part of the intended atmosphere, communal mythopoesis may be structurally less central.

## Missing Mechanisms: When Difficulty Alone Fails

Several difficult games did not generate the same sustained engagement, illustrating the consequences of missing or poorly implemented mechanisms.

Lords of the Fallen (Deck13 Interactive & CI Games, 2014) illustrates compromised cultivation. While it adopted familiar Souls mechanics, some player reviews described inconsistent hit detection and 'unfair' enemy behavior (see Metacritic n.d.). If death is perceived as arbitrary rather than pedagogical, the growth mindset underlying cultivation is harder to sustain. In such cases, aspirational deferment may weaken, and community discourse may center more on irritation than on future mastery.

The Surge (Deck13 Interactive, 2017) improved on this with competent mechanics, achieving moderate cultivation and deferment through its sci-fi progression. However, its more explicit narrative leaves less room for interpretive labor. That makes the game less a mythopoetic touchstone for the genre than a solid mechanical variant.

| Game | Ludic Cultivation | Aspirational Deferment | Communal Mythopoesis | Outcome |
|---|---|---|---|---|
| Elden Ring | Strong but permissive  Fair and varied, but open-world routing can | **Strong**  Long horizon, clear growth | **Strong**  Massive interpretive labor, cultural touchstones | **Ordeal Pleasure**  High engagement, cultural phenomenon |
| | bypass cultivation | | | |
| Hollow Knight | **Strong**<br><br>Precise, fair, challenging | **Strong**<br><br>Clear progression, "Pantheons" | **Less Difficulty-Coupled**<br><br>Rich lore but less structurally tied to difficulty; substantial but smaller-scale interpretive community | **High Engagement**<br><br>Critical success, distinct from "Souls" phenomenon |
| Lords of the Fallen (2014) | **Compromised**<br><br>Inconsistent, "unfair" feel | **Unsupported**<br><br>Frustration without growth | **Minimal**<br><br>Utilitarian strategy only | **Frustration**<br><br>Mixed reception, limited community |
| The Surge | **Moderate**<br><br>Competent but mechanical | **Moderate**<br><br>Clear goals, sci-fi setting | **Limited**<br><br>Exposition-heavy, low ambiguity | **Niche Success**<br><br>Solid game, limited cultural footprint |

**Table 1:** Comparative Summary of Mechanism Integration of example titles.

Note: Mechanism ratings are heuristic interpretations based on design analysis and secondary reception data, not standardized measurements.

Taken together, these cases suggest that large-scale interpretive labor is not always necessary for challenge-based engagement, but it is central to the fuller, community-wide form of ordeal pleasure theorized here.

The comparative analysis therefore points to ordeal pleasure as architecturally complex. Designers cannot assume that adding difficulty or ambiguity in isolation will reproduce the Souls-like pattern. What matters is how cultivation, deferment, and communal meaning-making are arranged, and how strongly they reinforce one another in a given game.

## DISCUSSION AND IMPLICATIONS

### Theoretical Contributions

The Ordeal Pleasure Framework makes three primary theoretical contributions:

1. Integration of temporal dynamics into motivation theory: SDT and flow theory largely describe present-moment states. Aspirational Deferment shows how present frustration becomes meaningful when players orient themselves toward future mastery. This helps explain the competence frustration paradox: players tolerate present frustration because they invest in future growth. The framework therefore argues that motivation theory needs a stronger account of temporality in intrinsically motivated play. We therefore frame Souls-like play as exposing a temporal motivation gap in existing play theory: standard models describe why challenge can be enjoyable in the present, but not well enough how players render prolonged present frustration meaningful through commitment to future mastery.

2. Specification of social mechanisms beyond generic relatedness: SDT's relatedness need is underspecified—any social connection satisfies it. Communal Mythopoesis specifies a particular form of relatedness: collaborative interpretive labor that constructs shared cultural meaning. This connects game studies to theories of interpretive communities (Fish 1980) and participatory culture (Jenkins 2006), showing how digital spaces enable collective hermeneutics. The framework suggests social motivation in games varies qualitatively—strategic cooperation, competitive interaction, and collaborative meaning-making are distinct modes—and recognizing this distinction matters for explaining why Souls communities differ in kind, not just degree, from other game communities. More specifically, OPF highlights interpretive relatedness: a form of social motivation grounded not merely in connection or cooperation, but in the collaborative production of shared meaning.

3. Synergy as explanatory framework: Most interdisciplinary integrations are additive. The OPF instead argues that cultivation generates material for mythopoesis, mythopoesis can scaffold cultivation, and deferment helps sustain both. This remains a theoretical and correlational claim; whether these mechanisms are causally interdependent rather than simply co-occurring remains for future empirical work.

The framework also suggests a broader research value in games. Because play takes place in controlled systems with observable behaviors, games provide useful settings for studying how people attach meaning to difficult experiences. If the fuller form of ordeal pleasure in games appears to rely on synergistic mechanisms, this generates hypotheses about ordeal experiences in other contexts (athletic training, artistic development, spiritual practice). An important open question is how cultural contexts shape the balance among these mechanisms. Initial work on Black Myth: Wukong suggests that culturally specific narrative substrates can reshape how communal meaning-making operates across global audiences (Fan 2025), though the relationship between cultural authenticity and the ordeal pleasure mechanisms proposed here remains to be investigated.

## Design Implications

The framework provides concrete guidance for designers attempting to create meaningfully difficult experiences:

Design for cultivation, not arbitrary punishment. Difficulty should implement desirable difficulties while remaining legible and learnable. Death should provide feedback that supports skill development. Inconsistent hit detection, opaque systems, or unreadable encounters undermine the perceived fairness on which cultivation depends. The 'git gud' philosophy works only when 'gud' is actually achievable through practice.

Architect temporal investment strategically. Long-form games (30+ hours) create conditions for Aspirational Deferment, but only if progression is visible. Players need evidence that investment pays off—early challenges becoming manageable, new capabilities enabling previously impossible actions. 'Come back later' affordances (open-world structure, multiple progression paths) give players agency in managing difficulty timing, preventing premature abandonment while maintaining challenge.

Give players something worth interpreting. Cryptic lore is not sufficient by itself, but rich narrative fragments, strategic ambiguity, and memorable cultural touchstones can invite collaborative interpretation. The challenge is to create ambiguity that rewards inference rather than simply obscuring information.

Design for community scaffolding. In-game communication systems (message systems, phantoms, summon signs) make community support visible and accessible. These mechanics serve both instrumental functions (strategic information) and psychological functions (normalizing failure, creating solidarity). Asynchronous multiplayer enables community support without demanding that players coordinate in real time.

Recognize design complexity. Games can succeed on one or two axes, as Hollow Knight shows, but the distinctive community intensity associated with Souls-like ordeal pleasure appears most clearly when cultivation, deferment, and mythopoesis reinforce one another.

Taken together, these principles also function as a diagnostic checklist for designers: when a difficult game fails to sustain player investment, the key question is not simply whether it is too hard, but which mechanism is weakest—cultivation, deferment, or mythopoesis—and what that weakness is costing the overall experience.

## Limitations, Boundary Conditions, and Future Directions

### Theoretical limitations

The framework is based primarily on successful cases, creating potential survivorship bias. We analyze players who persisted and engaged with the community, potentially overlooking those who disengaged under the same conditions. This survivorship problem is visible even in platform-wide progression data: as of March 2026, Steam global achievement data indicates that only a minority of Dark Souls III players reach the ending, while a somewhat larger but still limited share reach late-game milestones (Valve Corporation n.d.). Our analysis focuses on players who engaged. We do not account for those who found the difficulty arbitrary and quit. These mechanisms likely function as a filter: they may create ordeal pleasure for players with sufficient frustration tolerance, rather than generating it universally. Consequently, while Petralito et al. (2017) provide powerful evidence that frustration and enjoyment can coexist, their data likely reflects this 'survivor' cohort; broader longitudinal evidence would be needed to know how far this pattern extends across the gaming population.

### Core vs. Symbiosis

A central open question remains: Is the 'Forging Ritual' truly a triad, or is Ludic Cultivation (the technique) the sole necessary engine, with Mythopoesis (the legend) merely a cultural byproduct? While our comparative analysis suggests that deferment and mythopoesis can substantially intensify the fuller, community-wide phenomenon, it remains possible that for some players the Anvil alone is sufficient. The 'Forging' metaphor describes an ideal type of high-agency engagement; we acknowledge a spectrum of player roles, ranging from the 'Smith' who actively forges their path to the 'Apprentice' who relies heavily on community templates and guides.

Falsifiable Hypotheses: To resolve the 'Core vs. Symbiosis' debate, we propose two testable hypotheses for future research:

- H1: In a controlled experiment, removing narrative ambiguity (weakening the Legend) while keeping difficulty constant will significantly reduce 'long-term persistence' (churn rate) but not 'short-term learning curves'. This would support a strong intensifying role for Mythopoesis in sustaining the ordeal.
- H2: Providing immediate extrinsic rewards (e.g., real-money value or leaderboards) for failure will undermine the 'Ordeal Pleasure' transformation. We predict this would be observable as a reduction in 'persistence', lower self-reported 'intrinsic enjoyment', and a shift in 'failure attribution' (from 'I need to learn' to 'I lost the reward').

*Boundary Conditions and Player Variance*

Communal Mythopoesis also raises a community formation paradox. The framework implies that players benefit from an existing community to achieve the fullest ordeal pleasure-yet the community itself requires prior player engagement to form. This is not a contradiction: early adopters built these communities before their amplifying effects could operate, and later players then benefited from the scaffolding they created. However, it does suggest that the mechanism may function more as an intensifier that compounds over time than as a necessary precondition for initial engagement. This matters for new games attempting to cultivate the phenomenon from scratch.

*Generalizability questions*

The framework is derived from action RPGs with specific design conventions. Whether it generalizes to other difficulty-based genres (roguelikes, bullet hell, competitive games) remains an open question. While we expect partial transfer to games like Celeste (Matt Makes Games 2018) or Spelunky (Mossmouth 2009), the specific mythopoetic dimension may be less central in genres with less narrative ambiguity.

*Future research directions*

Longitudinal studies tracking individual players through extended engagement could map how mechanisms develop and interact over time. Experimental manipulation of lore crypticity or checkpoint spacing could test specific mechanism contributions and establish causality. Cross-cultural comparative studies could identify how cultural context shapes which mechanism players engage most strongly. Neuroimaging during play could investigate the frustration-satisfaction transformation. Systematic analysis of additional failed Souls-likes could further validate the framework's predictive power.

Future research could also investigate how the interaction of the three mechanisms changes as players move from relatively self-directed play to heavier reliance on guides and community scaffolding. Empirical studies could distinguish between seeking narrative meaning and seeking strategy optimization within communities, and trace how each relates to long-term persistence and intrinsic motivation.

## CONCLUSION

The puzzle this paper addresses is not why players merely endure repeated failure, but why some difficult games transform that failure into a durable source of meaning. The Ordeal Pleasure Framework argues that Souls-like games do so through the interaction of Ludic Cultivation, Aspirational Deferment, and Communal Mythopoesis.

The paper's central claim is therefore comparative rather than absolute: meaningful difficulty can exist without this exact alignment, but the Souls-like configuration appears to intensify and stabilize ordeal pleasure when these three mechanisms reinforce one another. OPF is not a universal theory of difficult play; it is a framework for identifying when difficult play becomes not merely challenging, but persistently meaningful, communally amplified, and intrinsically valued. Beyond Souls-likes, the framework offers a comparative lens for analyzing other difficult games and a hypothesis-generating scaffold for future empirical work on motivation, learning, and collective meaning-making.

The player's ordeal is a forging ritual: both smith and metal, emerging from the fire not broken but tempered. The ordeal is not what blocks pleasure. It is what forges it.